\documentclass[twocolumn,showpacs,amsmath,amssymb,prl,floatfix,superscriptaddress]{revtex4}

\usepackage{graphicx}
\usepackage{dcolumn}
\usepackage{bm}
\usepackage{bbm}
\usepackage{multirow,amssymb,amsbsy,amsmath}
\usepackage{stmaryrd}
\usepackage{color}

\newcommand{\ddim}{\udelta\kern0.1em}

\newcommand{\beikonst}[2]{\left( #1 \right)_{\kern-0.2em #2}}
\newcommand{\tr}[2][]{\text{Tr}_{#1}\left\{#2\right\}}

\newcommand*{\ket}[1]{\mathopen{|}#1\mathclose{\rangle}}

\newcommand{\dop}{{\rho}}

\newcommand{\s}{{\sigma}}

\newcommand{\ketbra}[1]{\mathopen{|}#1\mathclose{\rangle}\hspace{-0.25em}\mathopen{\langle}#1\mathclose{|}}

\begin{document}


%
%

\title{Long-range quantum gates using dipolar crystals}%

\author{Hendrik Weimer}%
\email{hweimer@cfa.harvard.edu}%
\affiliation{Physics Department, Harvard University, 17 Oxford Street, Cambridge, MA 02138, USA} %
\affiliation{ITAMP, Harvard-Smithsonian Center for Astrophysics, 60 Garden Street, Cambridge, MA 02138, USA}%
\author{Norman Y. Yao} %
\affiliation{Physics Department, Harvard University, 17 Oxford Street, Cambridge, MA 02138, USA} %
\author{Chris R. Laumann} %
\affiliation{Physics Department, Harvard University, 17 Oxford Street, Cambridge, MA 02138, USA} 
\affiliation{ITAMP, Harvard-Smithsonian Center for Astrophysics, 60 Garden Street, Cambridge, MA 02138, USA}%
\author{Mikhail D. Lukin} %
\affiliation{Physics Department, Harvard University, 17 Oxford Street, Cambridge, MA 02138, USA}

\date{\today}%

\begin{abstract}

We propose the use of dipolar spin chains to enable long-range quantum
logic between distant qubits. In our approach, an effective
interaction between remote qubits is achieved by adiabatically
following the ground state of the dipolar chain across the
paramagnet to crystal phase transition.  We demonstrate that the
proposed quantum gate is particularly robust against disorder and
derive scaling relations, showing that high-fidelity qubit
coupling is possible in the presence of realistic
imperfections. Possible experimental implementations in systems
ranging from ultracold Rydberg atoms to arrays of Nitrogen-Vacancy
defect centers in diamond are discussed.

\end{abstract}


\pacs{03.67.-a, 05.30.Rt, 76.30.Mi, 32.80.Ee}
\maketitle

The ability to carry out quantum gates between spatially remote qubits
forms a crucial component of quantum information processing
\cite{DiVincenzo2000}. Theoretical and experimental work addressing
this challenge has largely been focused upon using photons
\cite{Cirac1997,Moehring2007,Togan2010}, spin chains
\cite{Bose2003,Christandl2004,Gualdi2008,Yao2011} and other hybrid
systems \cite{Craig2004,Rikitake2005,Rabl2010} as quantum buses, which
mediate long-range quantum information transfer. In these approaches,
this transfer is achieved by either encoding the information in a
traveling wavepacket
\cite{Cirac1997,Moehring2007,Togan2010,Bose2003,Christandl2004,Gualdi2008},
or by coupling the remote qubits to a shared spatially extended mode
\cite{Yao2011,Craig2004,Rikitake2005,Rabl2010}.  In this Letter, we
propose a novel approach to this outstanding problem and demonstrate
that adiabatic driving of a dipolar spin system across a quantum phase
transition can be used to implement a robust controlled-phase (CP)
gate.

Our approach is applicable to dipolar spin systems
\cite{Rabl2007,Weimer2008a,Pohl2010,Schachenmayer2010,vanBijnen2011,Gorshkov2011},
composed, for example, of ultracold atoms and molecules, or
solid-state spin ensembles, where natural imperfections invariably
lead to disorder.  E.g., for spin qubits associated with
Nitrogen-Vacancy (NV) centers in diamond \cite{Childress2006,
Dutt2007}, the need for both long-range and disorder-robust quantum
gates is especially evident.  Despite room temperature coherence times
of $\sim 10\,\mathrm{ms}$, the weakness of magnetic dipole-dipole
interactions limits NV spacing to $\sim 10\,\mathrm{nm}$ for effective
two-qubit gates.  Even recently demonstrated sub-wavelength techniques
\cite{Maurer2010} cannot address individual NV qubits at such small
separations.  Moreover, any solid-state quantum bus designed to
mediate longer ranged interactions will suffer from positional
disorder due to the difficulty of precise nanoscale implantation.

\begin{figure}[t!b!h!]
  \includegraphics[width=0.9\linewidth]{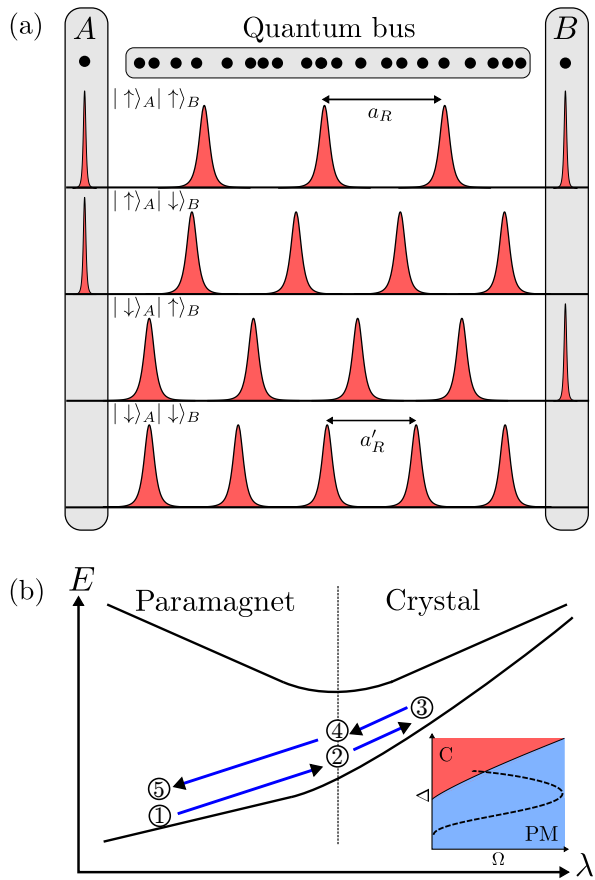}

  \caption{Setup for the proposed gate. (a) Depending on the state of
    the qubits $A$ and $B$, the quantum bus (atoms shown as black
    dots) in the crystalline phase will possess a different ground
    state, where the distance between two excitations $a_R$ is
    altered. This intuitively corresponds to a
    boundary-condition-dependent compression of the crystal
    ($a_R'<a_R$). (b) Ground state and first excited state during the
    control sequence. Initially, the quantum bus is prepared in the
    paramagnet (1). Then, the control parameter $\lambda$ is increased
    adiabatically (2), driving the system across the phase
    transition. Once in the crystal the system evolves freely and
    picks up a phase shift depending on the qubit states (3). After
    the adiabatic process is reversed (4), the quantum bus is
    disentangled from the qubits (5). The inset shows the phase
    diagram of the system, with the dashed line depicting the control
    profile dependent on the Rabi frequency $\Omega(t)$ and the
    detuning $\Delta(t)$.}

  \label{fig:intro}
\end{figure}

Here, we explore a possible solution to creating a quantum
bus within a disordered system. The key element underlying our
proposal is a phenomenon discussed in the context of Rydberg atoms as
the blockade effect \cite{Jaksch2000,Lukin2001,
Urban2009,Gaetan2009}. The simultaneous driving of two
excitations within length scales shorter than the blockade radius is
forbidden as strong interactions shift the doubly-excited state away
from resonance. Hence, within the blockade radius, the underlying
spatial distribution of the sites is largely irrelevant and the
arising many-body ground state washes over the effects of disorder and
can lead to the formation of crystalline structures
\cite{Weimer2008a}.

The dynamic crystal formation
\cite{Weimer2008a,Pohl2010,Schachenmayer2010,vanBijnen2011} underlying
our protocol is schematically illustrated in Fig. 1.  We consider two
qubits $A$ and $B$, coupled to the ends of a one-dimensional (1D)
quantum bus of two state atoms with one electronic ground state and
one Rydberg state. The quantum bus is initially prepared in its
ground state, containing no Rydberg excitations. Then, the bus
is adiabatically driven into the crystal regime. The resulting
many-body state has an energy which depends on the boundary conditions
set by the state of the qubits.  Intuitively, this dependence results
from a compression of the crystal, and hence, a decrease in the
distance between two Rydberg excitations, $a_R$, when the boundary qubits are
not excited. Under free evolution, this energy difference is
translated into a phase difference, which entangles the qubits. After
reversing the adiabatic step, the quantum bus returns to its initial
state while the qubits remain entangled.

To be specific, we consider an ensemble of strongly interacting
two-state systems described by the Hamiltonian
\begin{equation}
    H = -\frac{\hbar\Delta}{2} \sum\limits_i \s_i^z + \frac{\hbar\Omega}{2}\sum\limits_i\s_i^x + \sum_{i<j} \frac{C_p}{|{\bf r}_i - {\bf r}_j|^p} P^{\uparrow}_i P^{\uparrow}_j,
\label{eq:H}
\end{equation}
where $\Delta$ represents the detuning, $\Omega$ is the Rabi frequency
and $p=3$ for dipolar interactions or $p=6$ for van der Waals
interactions.  The interaction strength is characterized by the
coefficient $C_p$ and involve projectors onto one of the states,
$P^{\uparrow}_i =\ketbra{\uparrow} =
(1+\sigma^z_i)/2$. As we will discuss further below,
the same Hamiltonian also applies to NV centers under appropriate
driving. Here, the electronic ground state is a spin triplet; thus,
the $m_s=0$ state corresponds to the
atomic ground state, while the $m_s=1$ state, which possesses a
magnetic dipole moment, corresponds to the excited Rydberg state
\cite{Childress2006, Dutt2007, Neumann2010,Yao2010}. 

An analogous Hamiltonian governs the interactions between the boundary qubits (A,B) and the 
quantum bus,
\begin{equation}
  H_{int} = \sum_{i} \frac{C_p}{|{\bf r}_A - {\bf r}_i|^p} P^{\uparrow}_A P^{\uparrow}_i + \frac{C_p}{|{\bf r}_B - {\bf r}_i|^p} P^{\uparrow}_B P^{\uparrow}_i .
\end{equation}
Since the interaction conserves $\s^z_A$ and $\s^z_B$, any entangling
operation between the qubits will be in the form of a CP
gate. 
We may drive the quantum bus independently of the qubits by ensuring that the
resonant splitting of the qubits differs from that of the mediating bus;
possible implementations will be discussed later. 

To derive general scaling properties, we now
consider a one-dimensional system containing $N$ two-state spins
within length $L$. The CP gate protocol consists of an adiabatic ramp
from the classical ground state into the crystal regime for time
$t_0$, hold for phase accumulation for time $t_\pi$ and reverse ramp
(another $t_0$), resulting in a total gate time $t_g = 2t_0+t_{\pi}$.
There are three factors which influence the asymptotic scaling of the
fidelity with system size: (i) the strength of the effective
interaction $E_{\textrm{int}}$ at the hold point, (ii) the minimum
energy gap $\Delta_g$ (across ramp and qubit sectors) protecting the
adiabatic evolution, and (iii) the strength of external decoherence
mechanisms.  The interaction energy between the qubits, which governs
the timescale of entanglement generation, is
\begin{equation}
  E_{\mathrm{int}} = E_{\uparrow\uparrow}-E_{\uparrow\downarrow}-E_{\downarrow\uparrow}+E_{\downarrow\downarrow},
\end{equation}
where $E_{\alpha\beta}$ refers to the energy of the many-body state
with the qubits in state $\ket{\alpha}_A\ket{\beta}_B$
(Fig.~\ref{fig:intro}).  Within the continuum limit of a classical
crystal, $E_{\mathrm{int}} \sim d^2/L$ for $d\ll a_R$, where
$d$ is the distance between the qubits and the ends of the quantum
bus. Owing to quantum fluctuations, the classical crystal cannot be
the true ground state, and the system is rather described in terms of
a Luttinger liquid \cite{Weimer2010a,Sela2011}.  However, such
corrections are important only in the limit of very large system sizes
($N\sim 10^9$ for typical parameters), where they lead to an algebraic
decay of the correlation functions \cite{Weimer2010a}.

To analyze the effects of a finite gap and decoherence, we consider
the contributions of each to the overall error of the controlled phase
gate, assuming they occur independently \cite{Lacour2007}.  While, in
the thermodynamic limit, the gap vanishes at the phase transition,
here, we consider finite system sizes where there always exists a
non-zero gap. For gapless phases such as the dipolar crystal however,
it is important to note that the gap may further decrease upon
entering the ordered phase. The qualitative effect of such a finite
gap is described within a Landau-Zener framework. Optimizing the form
of the Landau-Zener sweep by introducing a nonlinearity results in an
improved error scaling with, $\varepsilon_{LZ} = \exp(- c\Delta_G t_0
/\hbar)$, where $c$ is a model-dependent numerical constant
\cite{Roland2002,Quan2010}.  In the case of the dipolar crystal,
$\Delta_G\sim 1/L$, due to the phononic nature of the excitations
\cite{Weimer2010a,Sela2011}; this results in an error which scales
according to the theoretical optimum given by the Lieb-Robinson bound
for the speed of information transfer \cite{Lieb1972}.

Next, we consider the effects of decoherence, noting that the induced
error is a monotonically increasing function dependent only on the
product of the decoherence rate $\gamma$ and the total gate time
$t_g$. In particular, we assume, $\varepsilon_d = 1-\exp[-(\gamma
t_g)^\delta]\approx(\gamma t_g)^\delta$, where $\delta$ depends on the
physical details of the decoherence process \cite{Maze2008}. The
highly entangled nature of the various many-body states depicted in
Fig.~\ref{fig:intro} implies that the effective decoherence rate must
scale with the system size, $\gamma = \gamma_0\frac{L}{L_0}$, where
$\gamma_0$ is the single particle decoherence rate and the length
scale $L_0$.  In the dipolar crystal, $L_0$ is approximately given by
the average distance between two excited spins; this is consistent
with intuition, as decoherence processes ought only be relevant at
sites where there exists an actual excitation.

To separate off the explicit system size dependence within
$\varepsilon_{LZ}$, we define $\alpha_0 =
cL\Delta_G/\hbar$. Combining the two error contributions and
conservatively plugging $t_g > t_0$ into the form for $\epsilon_{LZ}$,
then yields
\begin{equation}
  \varepsilon_T = \exp\left(-\frac{\alpha_0}{L} t_g\right) +\left(\gamma_0 \frac{L}{L_0} t_g\right)^\delta.
  \label{eq:err}
\end{equation}
By minimizing the total error, we obtain an optimal gate time, $t_g^{\mathrm{opt}}= \delta L\log[L_0\alpha_0/(L^2\gamma_0)]/\alpha_0$, with corresponding error,
\begin{equation}
  \varepsilon_T = L^{2\delta} \left(\delta\frac{\gamma_0}{L_0\alpha_0}\log\frac{L_0\alpha_0}{L^2\gamma_0}\right)^\delta.
\end{equation}
Thus, our protocol exhibits a scaling analogous to a quantum gate
based on a microscopic interaction with energy $C_2/L^2$, which has an
error given by $\varepsilon =
L^{2\delta}(2\pi\gamma_0/C_2)^\delta$. The precise values of
$\alpha_0$ and $L_0$ can be derived from numerical studies, as we will
show in the following.

\emph{Simulations.---} In these numerical simulations, we both verify
our general scaling arguments and demonstrate the ability to achieve
superior gate fidelities in comparison to bare microscopic
interactions.  We begin by considering a chain of $N$ equidistant or
uniformly randomly distributed particles with average interparticle spacing
$a$; the details of the numerical simulation method are described in
\cite{Weimer2008a}.  Initially, the qubits are prepared in the state
$\ket{\psi}_{A,B} = (\ket{\uparrow}_A+\ket{\downarrow}_A)\otimes
(\ket{\uparrow}_B+\ket{\downarrow}_B)/2$, while the spin chain is
fully polarized, $\ket{\psi}_{SC} = \prod_i \ket{\downarrow}_i$. This
initial qubit state consitutes a worst-case scenario leading to a
minimum value of the fidelity for relevant decoherence models,
including pure dephasing. Using a different initial state for the qubits
can lead to a significantly higher fidelity. To drive the system
across its phase transition, external control fields are then varied
according to
\begin{eqnarray}
  \Omega(t) &=& \Omega_0 \sin^2\left(\frac{8t/t_0}{1+16t^2/t_0^2}\right)\\
  \Delta(t) &=& \Delta_0 [1-5\exp(-4t/t_0)].
\end{eqnarray}
Here, $\Omega_0$ is chosen such that the endpoint of the ramp lies
just within the crystalline phase, see Fig.~\ref{fig:intro}. While the
proposed ramp profile features the requisite nonlinearity, its details
have yet to be optimized; therefore, with optimal control theory, it
may be possible to further enhance the achievable gate fidelities
\cite{Khaneja2001}. At $t=t_0$, the system freely evolves for a time,
$t_\pi = \pi\hbar/E_{\mathrm{int}}$, in order to allow the effective
interaction to generate a phase gate between the qubits.  Following
this period of free evolution, the adiabatic ramp is then reversed. In
addition to naturally following the reversed profile, an alternate
implementation can also be achieved by a complete reversal of
Hamiltonian dynamics upon switching $H$ to $-H$. The ability to change
the sign of the interaction depends on the physical implementation; in
the case of Rydberg atoms, this can be achieved by transferring the
population from a repulsive state to an attractive state or by
changing an external electric field \cite{Carroll2004}. We find that
both protocols give nearly identical results and focus on the latter,
as it simplifies the numerical analysis.  Since our procedure
implements a controlled phase gate up to local rotations, the fidelity
of the proposed gate is then given by the disentanglement fidelity
between the qubits and the chain, $F_{LZ} =\sqrt{\tr{\dop_{AB}^2}}$,
where $\dop_{AB}$ is the qubits' reduced density matrix.

\begin{figure}[tb!]
\includegraphics{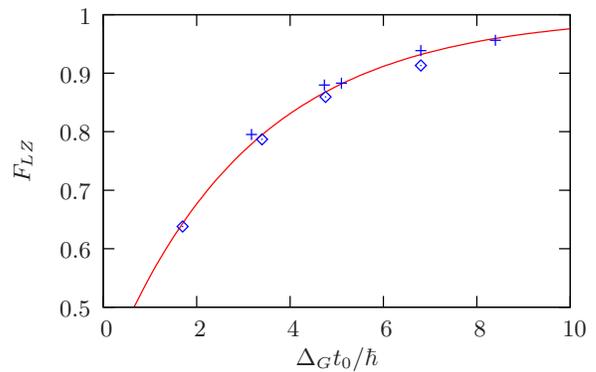}
\caption{Numerical simulation results for the gate fidelity $F_{LZ}$
  due to non-adiabatic transitions depending on the product of the gap
  $\Delta_G$ and duration of the adiabatic step $t_0$ for equidistant
  ($N=34$, diamonds) and disordered ($N=15$, crosses)
  configurations. The gap has been varied by changing $\Omega_0$ while
  all other parameters are held fixed. The solid line is an
  exponential fit to the data. ($p=3$, $C_3 = 100\,\hbar\Omega_0 a^3$,
  $\Delta_0 = 2.3\,\Omega_0$, $d = 3\,a$).}
\label{fig:num}
\end{figure}
In the absence of decoherence, we expect the errors to be
characterized by a Landau-Zener exponential and this is indeed
revealed by the simulations, as shown in Fig. 2; the fidelity is
well-characterized by $F=1-b\exp(-c \Delta_G t_0/\hbar)$, where $b$
and $c$ are numerical fit parameters. Combining these numerics with
the additional errors associated with decoherence ($\varepsilon_d$)
yields an overall fidelity,
\begin{equation}
  F_T = \frac{1}{2}[1-be^{-c \Delta_G t_0/\hbar}]\left [1+e^{-\left(\gamma_0 \frac{L}{L_0} \left[2t_0 + \frac{\pi \hbar}{E_{int}}\right]\right)^\delta}\right ].
\label{eq:f}
\end{equation}
Note that for near-perfect gates this expression
is equivalent to Eq.~(\ref{eq:err}). As previously discussed, there
now exists a maximum fidelity, which is achieved by an optimal ramp
time, $t_0$ that is a function of only the effective interaction
strength, $E_{int}$, and the gap, $\Delta_G$.

\begin{figure}[t!b!]
\includegraphics{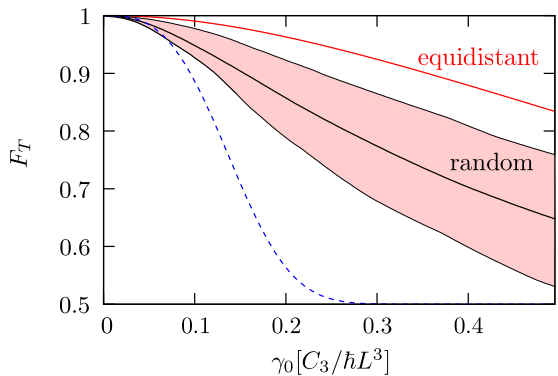}
\caption{Dependence of the maximum fidelity of the proposed quantum
  gate on the single particle decoherence rate $\gamma_0$ with
  parameters taken from the numerical simulation for a system size of
  $L$ and a spin-echo suppressed decoherence rate with $\delta \approx
  3$ . The solid red line is the fidelity in the equidistant case with
  $N=34$ particles, while the shaded areas correspond to 90\%
  confidence intervals for a disordered situation ($N=51$). The dashed
  line indicates the fidelity that can be achieved using the bare
  dipolar interaction between the qubits.}
\label{fig:deco}
\end{figure}

Crucially, Eq.~(\ref{eq:f}) now allows us to investigate the
consequences of a disordered interparticle spacing. For a 1D dipolar
crystal, it is known that the crystal spacing, $a_R =
[\zeta(p)(p+1)C_p/\Delta]^{1/p}$, is essentially independent of the
spacing between individual particles, suggesting that the crystalline
phase may be robust against effects of disorder \cite{Weimer2010a}. To
evalutate the gate fidelity (\ref{eq:f}), we numerically
determine the gap, $\Delta_G$, and the interaction energy,
$E_{\mathrm{int}}$, for 100 different uniformly distributed random
configurations for the parameters as in Fig.~\ref{fig:num} ($b=0.62$,
$c=0.32$). By extracting $L_0=L/\sum_i\langle
P_i^\uparrow\rangle=2.0\,(C_3/\Omega_0)^{1/3}$ from the average
density of excitations and employing Eq.~(\ref{eq:f}), we then
calculate the optimum adiabatic ramp time $t_0$ and hence, the maximum
gate fidelity. The result is shown in
Fig.~\ref{fig:deco}. Significantly, \emph{even} in the presence of
disorder, the fidelity of our phase gate is higher than that which can
be achieved via microscopic dipolar interactions.

\emph{Experimental Realization.---} In the case of Rydberg atoms, it
is possible to achieve independent addressing of the boundary qubits
by choosing differing hyperfine levels for the qubit and quantum bus
atoms.  Our proposed protocol can be realized through either van der
Waals interactions ($p=6$) between $S$ states or through dipolar
interactions ($p=3$) between states within an electric-field induced
Stark fan \cite{Pupillo2010}.  Let us focus on the latter case and
consider a Rydberg state with principle quantum number $n = 43$ and
decoherence rate $\gamma_0 = 10\,\mathrm{kHz}$ ($\delta = 1$),
neglecting the effects of atomic motion, which is well justified at
temperatures on the order of $100\,\mathrm{nK}$. In order to achieve a
gate fidelity, $F=0.9$, the requisite laser parameters are given by
$\Omega_0 = 2\pi\times 8\,\mathrm{MHz}$, $\Delta_0 = 2\pi \times
17\,\mathrm{MHz}$, and the corresponding interaction strength is given
by $C_3a^{-3} = 2\pi\times 800\,\mathrm{MHz}$. Such an interaction
strength can be achieved in a Rydberg atom cloud with an average
interparticle spacing of $a \approx 1\mathrm{\mu m}$, leading to
$L\approx 40\,\mathrm{\mu m}$; we note that these parameters are
compatible with present experimental techniques
\cite{Saffman2010}. For van der Waals interactions, the enhancement of
the phase gate fidelity is even more pronounced.

It is also possible to implement our protocol using a solid-state room
temperature setup based upon NV defect centers in diamond
\cite{Childress2006, Dutt2007}. To realize the Hamiltonian
(\ref{eq:H}) using a quasi-1D chain of NV centers
\cite{Toyli2010,Spinicelli2011,Hausmann2011}, we work with the
$m_s=0,1$ electronic spin states with crystal field splitting $\approx
2.8\,\mathrm{GHz}$.  Independent addressing of the qubits and quantum
bus can be achieved by using a different Nitrogen isotope ($^{15}$N)
for the qubits and the bus ($^{14}$N); the hyperfine coupling between
the NV electronic spin and the Nitrogen nuclear spin is isotope
dependent, with $A_{\parallel}^{N15} \approx 3.03\,\mathrm{MHz}$ and
$A_{\parallel}^{N14} \approx -2.14\,\mathrm{MHz}$
\cite{Felton2009}. This difference ensures that the microwave driving
of the quantum bus is off-resonant with the qubit splitting, allowing
for independent initialization and control. Alternatively, the
boundary qubits may be taken as bare Nitrogen impurities manipulated
by a nearby NV center. The Nitrogen electron spin functions as the
boundary spin for the protocol and features resonance frequencies
detuned by $\mathrm{GHz}$, which potentially allows for stronger
driving during the gate procedure.

While recent experiments have demonstrated optical initialization of
the $m_s =0$ state with approximately $92$-$95$\% fidelity, it may be
possible to further enhance this initialization by exploiting the
neutral NV$^0$ charge state. In particular, recent work
\cite{Waldherr2011} has shown that red laser excitation can transfer
nearly $100$\% of the population to one $m_s$ sublevel of the
spin-$1/2$ NV$^0$ electronic spin. This enables us to effectively
polarize the NV nuclear spins of the chain via cycles of microwave
and red laser driving. Finally, after returning the defect to the
NV$^-$ charge state, a SWAP gate transfers the polarization from the
nuclear spins back to the spin-$1$ electronic spins of the NV$^-$.
Such enhanced initialization may prove beneficial for other NV-based
quantum computing architectures \cite{Yao2010}.

Magnetic field fluctuations (e.g., from a nuclear spin bath), which
give rise to $T_2^*$ dephasing processes can effectively be suppressed
by stroboscopically switching the system between the $m_s=\pm 1$
states. In the resulting dynamics, the electronic spin is then
decoupled from the environment and coherence times up to the spin
relaxation time $T_1$ can be achieved \cite{deLange2010}. This
procedure also leads to the suppression of undesired flip-flop
couplings between the NV centers. Assuming $\gamma_0 =
100\,\mathrm{Hz}$ \cite{Balasubramanian2009}, a Rabi frequency
$\Omega_0 = 2\pi\times 62\,\mathrm{kHz}$, a detuning $\Delta_0 =
2\pi\times 130\,\mathrm{kHz}$, and an average NV spacing of $a =
2\,\mathrm{nm}$, according to the results shown in
Fig.~\ref{fig:deco}, we can achieve gate times $t_g \approx
500\,\mathrm{\mu s}$ and fidelities of $F=0.98$ in the equidistant
case, and $F=0.93\pm 0.04$ for disordered configurations over a
distance of $L=74\,\mathrm{nm}$. We stress that such qubit distances
are compatible with the individual optical addressing and readout of
NV qubits using experimentally demonstrated sub-wavelength techniques
\cite{Maurer2010, Rittweger2009}.

In summary, we have shown that a robust long-range quantum gate can be
created using dipolar spin chains. We have discussed possible
experimental realizations with Rydberg atoms or NV centers and
emphasize that the proposed long-range gate can tolerate disorder. At
the same time, the proposed gate is not limited to the case of dipolar
crystals; indeed, one can implement our protocol within the transverse
Ising model, wherein even a nearest-neighbor interaction can be used
to create an effective $1/L^2$ power law interaction. Finally, our
proposal is also a first step towards studying quantum many-body
physics with NV centers.

We acknowledge fruitful discussions with P. Zoller, L. Huijse, and
A. Chandran. This work was supported by the National Science
Foundation through a grant for the Institute for Theoretical Atomic,
Molecular and Optical Physics at Harvard University and Smithsonian
Astrophysical Observatory, a fellowship within the Postdoc Program of
the German Academic Exchange Service (DAAD), the DOE (FG02-
97ER25308), the Lawrence Golub Fellowship, CUA, NSF, DARPA, AFOSR
MURI, and the Packard Foundation.


\begin{thebibliography}{10}

\bibitem{DiVincenzo2000}
D.~P. DiVincenzo,
\newblock Fortschr. Phys. {\bf 48}, 771 (2000).

\bibitem{Cirac1997}
J.~I. Cirac et~al.,
\newblock Phys. Rev. Lett. {\bf 78}, 3221 (1997).

\bibitem{Moehring2007}
D.~L. {Moehring} et~al.,
\newblock \nat {\bf 449}, 68 (2007).

\bibitem{Togan2010}
E.~{Togan} et~al.,
\newblock Nature {\bf 466}, 730 (2010).

\bibitem{Bose2003}
S.~Bose,
\newblock Phys. Rev. Lett. {\bf 91}, 207901 (2003).

\bibitem{Christandl2004}
M.~Christandl et~al.,
\newblock Phys. Rev. Lett. {\bf 92}, 187902 (2004).

\bibitem{Gualdi2008}
G.~Gualdi et~al.,
\newblock Phys. Rev. A {\bf 78}, 022325 (2008).

\bibitem{Yao2011}
N.~Y. Yao et~al.,
\newblock Phys. Rev. Lett. {\bf 106}, 040505 (2011).

\bibitem{Craig2004}
N.~J. Craig et~al.,
\newblock Science {\bf 304}, 565 (2004).

\bibitem{Rikitake2005}
Y.~Rikitake and H.~Imamura,
\newblock Phys. Rev. B {\bf 72}, 033308 (2005).

\bibitem{Rabl2010}
P.~{Rabl} et~al.,
\newblock Nature Phys. {\bf 6}, 602 (2010).

\bibitem{Rabl2007}
P.~Rabl and P.~Zoller,
\newblock Phys. Rev. A {\bf 76}, 042308 (2007).

\bibitem{Weimer2008a}
H.~Weimer et~al.,
\newblock Phys. Rev. Lett. {\bf 101}, 250601 (2008).

\bibitem{Pohl2010}
T.~Pohl, E.~Demler, and M.~D. Lukin,
\newblock Phys. Rev. Lett. {\bf 104}, 043002 (2010).

\bibitem{Schachenmayer2010}
J.~{Schachenmayer} et~al.,
\newblock New J. Phys {\bf 12}, 103044 (2010).

\bibitem{vanBijnen2011}
R.~M.~W. van Bijnen et~al.,
\newblock J. Phys. B {\bf 44}, 184008 (2011).

\bibitem{Gorshkov2011}
A.~V. Gorshkov et~al.,
\newblock Phys. Rev. Lett. {\bf 107}, 115301 (2011).

\bibitem{Childress2006}
L.~Childress et~al.,
\newblock Science {\bf 314}, 281 (2006).

\bibitem{Dutt2007}
M.~V.~G. {Dutt} et~al.,
\newblock Science {\bf 316}, 1312 (2007).

\bibitem{Maurer2010}
P.~C. {Maurer} et~al.,
\newblock Nature Phys. {\bf 6}, 912 (2010).

\bibitem{Jaksch2000}
D.~Jaksch et~al.,
\newblock Phys. Rev. Lett. {\bf 85}, 2208 (2000).

\bibitem{Lukin2001}
M.~D. Lukin et~al.,
\newblock Phys. Rev. Lett. {\bf 87}, 037901 (2001).

\bibitem{Urban2009}
E.~Urban et~al.,
\newblock Nature Phys. {\bf 5}, 110 (2009).

\bibitem{Gaetan2009}
A.~Ga\"etan et~al.,
\newblock Nature Phys. {\bf 5}, 115 (2009).

\bibitem{Neumann2010}
P.~{Neumann} et~al.,
\newblock Nature Phys. {\bf 6}, 249 (2010).

\bibitem{Yao2010}
N.~Y. {Yao} et~al.,
\newblock arXiv:1012.2864  (2010).

\bibitem{Weimer2010a}
H.~Weimer and H.~P. B\"uchler,
\newblock Phys. Rev. Lett. {\bf 105}, 230403 (2010).

\bibitem{Sela2011}
E.~Sela, M.~Punk, and M.~Garst,
\newblock Phys. Rev. B {\bf 84}, 085434 (2011).

\bibitem{Lacour2007}
X.~Lacour et~al.,
\newblock Phys. Rev. A {\bf 75}, 033417 (2007).

\bibitem{Roland2002}
J.~Roland and N.~J. Cerf,
\newblock Phys. Rev. A {\bf 65}, 042308 (2002).

\bibitem{Quan2010}
H.~T. Quan and W.~H. Zurek,
\newblock New J. Phys. {\bf 12}, 093025 (2010).

\bibitem{Lieb1972}
E.~H. Lieb and D.~W. Robinson,
\newblock Commun. Math. Phys. {\bf 28}, 251 (1972).

\bibitem{Maze2008}
J.~R. Maze, J.~M. Taylor, and M.~D. Lukin,
\newblock Phys. Rev. B {\bf 78}, 094303 (2008).

\bibitem{Khaneja2001}
N.~Khaneja, R.~Brockett, and S.~J. Glaser,
\newblock Phys. Rev. A {\bf 63}, 032308 (2001).

\bibitem{Carroll2004}
T.~J. Carroll et~al.,
\newblock Phys. Rev. Lett. {\bf 93}, 153001 (2004).

\bibitem{Pupillo2010}
G.~Pupillo et~al.,
\newblock Phys. Rev. Lett. {\bf 104}, 223002 (2010).

\bibitem{Saffman2010}
M.~Saffman, T.~G. Walker, and K.~M{\o}lmer,
\newblock Rev. Mod. Phys. {\bf 82}, 2313 (2010).

\bibitem{Toyli2010}
D.~M. Toyli et~al.,
\newblock Nano Lett. {\bf 10}, 3168 (2010).

\bibitem{Spinicelli2011}
P.~Spinicelli et~al.,
\newblock New J. Phys. {\bf 13}, 025014 (2011).

\bibitem{Hausmann2011}
B.~J.~M. Hausmann et~al.,
\newblock New J. Phys. {\bf 13}, 045004 (2011).

\bibitem{Felton2009}
S.~Felton et~al.,
\newblock Phys. Rev. B {\bf 79}, 075203 (2009).

\bibitem{Waldherr2011}
G.~Waldherr et~al.,
\newblock Phys. Rev. Lett. {\bf 106}, 157601 (2011).

\bibitem{deLange2010}
G.~{de Lange} et~al.,
\newblock Science {\bf 330}, 60 (2010).

\bibitem{Balasubramanian2009}
G.~{Balasubramanian} et~al.,
\newblock Nature Mater. {\bf 8}, 383 (2009).

\bibitem{Rittweger2009}
E.~{Rittweger} et~al.,
\newblock Nature Photon. {\bf 3}, 144 (2009).

\end{thebibliography}

\end{document}